\documentclass[prl,aps,twocolumn, a4paper,longbibliography,superscriptaddress]{revtex4-2}

\pdfoutput=1

\date{\today}

\newcommand{\mum}{\upmu{\rm m}}
\newcommand{\kB}{k_{\textrm{B}}}

\newcommand{\kD}{k_{\textrm{D}}}
\newcommand{\kF}{k_{\textrm{F}}}
\newcommand{\UD}{U_{\textrm{D}}}
\newcommand{\tD}{t_{\textrm{D}}}

\newcommand{\Us}{U_{\textrm{s}}}

\newcommand{\Ein}{\epsilon_{\textrm{in}}}

\def\appropto{%
  \def\p{%
    \setbox0=\vbox{\hbox{$\propto$}}%
    \ht0=0.6ex \box0 }%
  \def\s{%
    \vbox{\hbox{$\sim$}}%
  }%
  \mathrel{\raisebox{0.7ex}{%
      \mbox{$\underset{\s}{\p}$}%
    }}%
}

\usepackage{float}

\usepackage{fourier}

\usepackage{color}
\usepackage{graphicx}
\usepackage{amsmath,amssymb}

\usepackage{times}
\usepackage{upgreek}
\usepackage{psfrag} 
\usepackage{latexsym} 
\usepackage{amstext}
\usepackage{amsxtra} 

\usepackage{lineno}

\usepackage{textcomp}
\usepackage{amsfonts}
\usepackage{graphicx}
\usepackage{bm}
\usepackage{color}
\usepackage{braket}
\usepackage{siunitx}
\usepackage[svgnames]{xcolor}
\usepackage[normalem]{ulem}

\definecolor{myColor}{rgb}{0.02,0.12,0.3}
\definecolor{myciteColor}{rgb}{0.39,0.7,0.89}
\usepackage[colorlinks=true,citecolor=myColor,linkcolor=myColor,urlcolor=myColor]{hyperref}
\definecolor{myChangesColor}{rgb}{0.1,0.1,0.8}

\def\be{\begin{equation}}
\def\ee{\end{equation}}

\makeatletter
\def\@fnsymbol#1{\ensuremath{\ifcase#1\or *\or \dagger\or \ddagger\or
   \mathsection\or \mathparagraph\or \|\or **\or \dagger\dagger
   \or \ddagger\ddagger \else\@ctrerr\fi}}
\makeatother

\begin{document}

\title{
An Equation of State for Turbulence in the Gross-Pitaevskii Model
}

\author{Gevorg~Martirosyan}
\email{gevorgmartirosyan97@gmail.com}
\affiliation{
Cavendish Laboratory, University of Cambridge, J. J. Thomson Avenue, Cambridge CB3 0HE, United Kingdom}
\author{Kazuya~Fujimoto}
\affiliation{Department of Physics, Institute of Science Tokyo, 2-12-1 Ookayama, Meguro-ku, Tokyo 152-8551, Japan}
\author{Nir~Navon}
\affiliation{
Department of Physics, Yale University, New Haven, Connecticut 06520, USA}
\affiliation{Yale Quantum Institute, Yale University, New Haven, Connecticut 06520, USA}

\begin{abstract}

{We report the numerical observation of a far-from-equilibrium equation of state (EOS) in the Gross-Pitaevskii (GP model. We first show that the momentum distribution of the turbulent cascade is well described by wave-turbulent kinetic theory in the appropriate limits. Calculating the energy and particle fluxes $\Pi_\varepsilon(k)$ and $\Pi_N(k)$, we show that the turbulent state possesses the hallmarks of a direct energy cascade. Building on this, we show that the GP model encodes a universal EOS in the form of a relationship between the turbulent cascade's momentum distribution amplitude $n_0$ and the energy flux $\epsilon$ in the steady state. We find that in our regime of `mixed' turbulence - where both vortices and waves play a significant role - $n_0\propto \epsilon^{0.67(2)}$, a result that is not captured by any existing theory of turbulence but that agrees with a recent experimental measurement for large energy fluxes. 
Finally, we find that the concept of quasi-static thermodynamic processes between equilibrium states extends to far-from-equilibrium steady states.
} 

\end{abstract}
\maketitle 

{\textit{Introduction.}} Equilibrium and near-equilibrium thermodynamics form conceptual cornerstones of physics, reducing the behavior of complex many-body systems to relations among a few macroscopic observables.  
Far from equilibrium, such a unifying framework is lacking, and unveiling universal features has become a major goal of modern physics.  
When local equilibrium holds, local thermodynamics and hydrodynamics provide a bridge to universal descriptions~\cite{Hohenberg:1977,Bray:2002,forster2018hydrodynamic}.  
Even without microscopic equilibrium, universal phenomena have been predicted and observed—ranging from thermalization of far-from-equilibrium states~\cite{Berges:2008,Erne:2018,Pruefer:2018,Glidden:2021,Huh:2024,Moreno:2025,Gazo:2025,Martirosyan:2025} to steady states characterized by state variables linked through far-from-equilibrium analogs of equations of state (EOS)~\cite{Takatori:2015,Ginot:2015,Dogra:2023}.

A remarkable example of steady states that remain locally far from equilibrium is matter-wave turbulence, sustained by energy injection and dissipation at distinct length scales.  
Such cascades have recently been realized in ultracold-atom systems~\cite{Navon:2016,Navon:2019,Zhao:2025}, where a far-from-equilibrium EOS was measured~\cite{Dogra:2023}.  
All tractable theoretical descriptions of these experiments rely on the classical-field Gross–Pitaevskii (GP) equation~\cite{Chantesana:2019,Sano:2022,Zhu:2023a,Fischer:2024}\footnote{This model is not limited to weakly interacting BECs; it also applies in the classical-field limit of highly occupied modes~\cite{Sinatra:2001,Davis:2002}.}.

Although the GP model has been studied for decades~\cite{fibich2015nonlinear,Pitaevskii:2016}, and its equilibrium and near-equilibrium behavior are well established, much of its far-from-equilibrium physics remains to be understood.  
It supports regimes such as Kolmogorov vortex turbulence~ \cite{Nore:1997,Kobayashi:2005,Tsubota:2017,kobayashi2021quantum,Tanogami:2021,Zhao:2025} and weak-wave turbulence (WWT)~\cite{Zakharov:1992,Nazarenko:2011}, which feature a far-from-equilibrium EOS relating the amplitude of a steady-state spectrum to an energy flux.  
However, these solutions do not describe the recently measured experimental EOS~\cite{Dogra:2023}, motivating a calculation of the EOS within the GP model in the same setting as the experiments.

In this Letter, we perform simulations of the GP model in the setting of the experiments~\cite{Dogra:2023} and show that the GP model possess a far-from-equilibrium EOS that lies outside any known paradigm of turbulence. The turbulent state obtained at long times is steady and we find that it is characterized by a momentum distribution quantitatively described by a universal prediction from WWT theory. This steady state possesses the hallmarks of a direct energy cascade, \emph{i.e.}\ that the dissipation-scale-independent energy flux is scale invariant, transporting energy from large to small length scales. 
However, the power-law scaling of the amplitude of the momentum distribution with the energy flux is starkly different from predictions of any theory of turbulence, but agrees well with the experimental EOS for large values of the energy flux.

\begin{figure}[t!]
\centerline{\includegraphics[width=1\columnwidth]{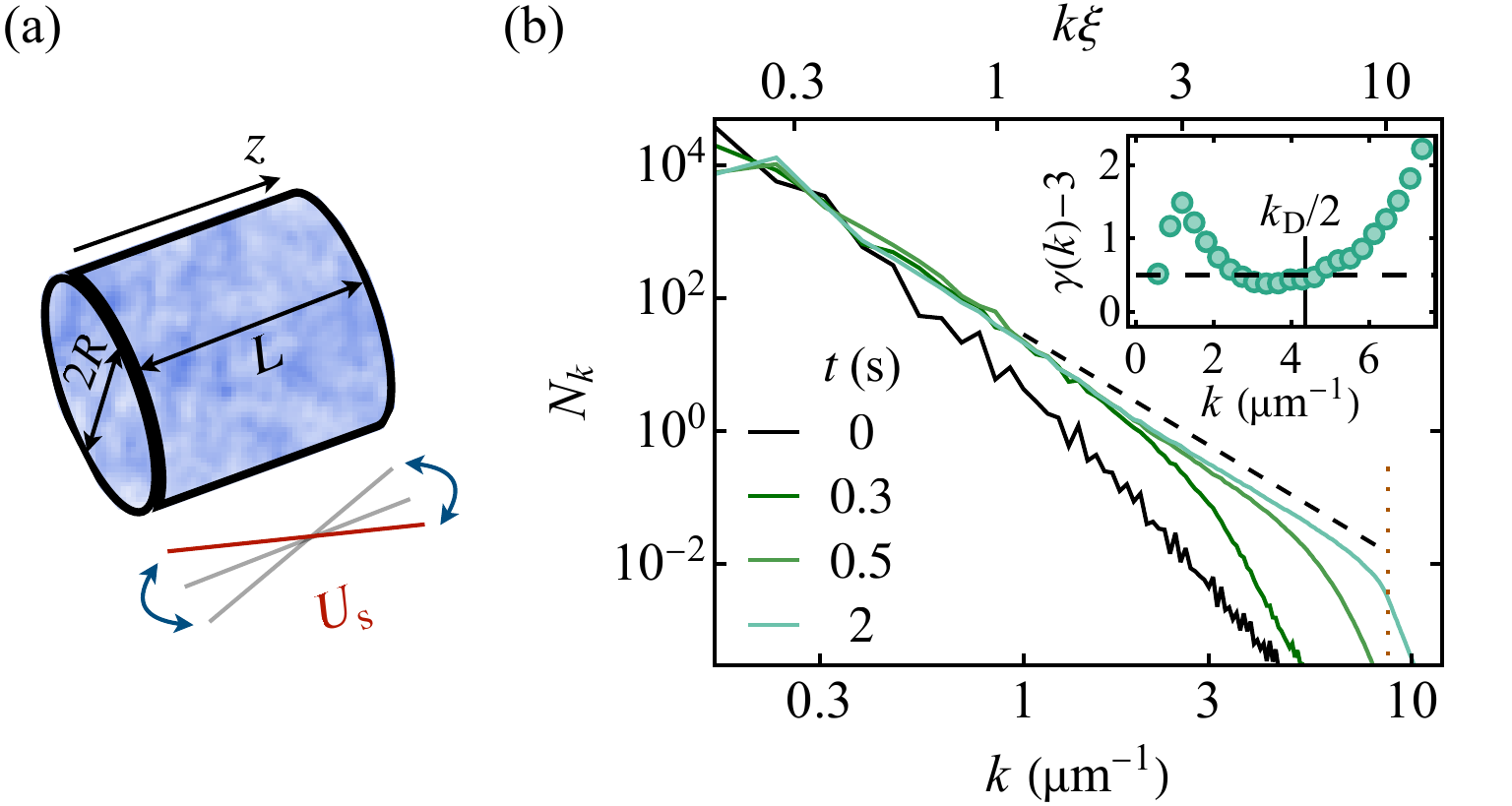}}
\caption{The direct turbulent cascade in the GP model. (a) Cartoon of the simulation geometry and the driving protocol. We use a cylindrical box trapping potential of length $L$ and radius $R$, and the energy is injected into the system by applying a time-periodic potential gradient $V_\text{drive}(\boldsymbol{r},t)=\Us  \sin(\omega t)z/L$ (see text for typical parameters).
(b) The build-up of the turbulent cascade; the mode occupation number $N_k$ is shown for various shaking times $t$. 
At long times, the system is in a steady state with $N_k\propto k^{-\gamma}$ (the dashed line corresponds to $\gamma=3.5$). The inset shows the cascade exponent $\gamma(k)\equiv-\textrm{d}\ln[n(k)]/\textrm{d}\ln{[k]}$, calculated from the continuous momentum distribution $n(k)$; the dashed line is $\gamma(k)=3.5$. Here, the simulation parameters are $L=50\,\mum$, $R=15\,\mum$, $\Us=1.0\zeta$, $\omega=2\pi\times10\,{\rm Hz}$, and $a=100a_0$ (corresponding to $\xi=1.2\,\mum$ and $\tau=10\,$ms). 
}
\label{fignk}
\end{figure}

\textit{The model.} Our study is based on the universal GP model 
\begin{equation}
\label{eq:GP0}
    i\hbar\frac{\partial \psi}{\partial t}=\left(-\frac{\hbar^2}{2m}\nabla^2+g|\psi|^2\right)\psi,
\end{equation}
for the field $\psi(\boldsymbol{r},t)$. This equation, also known as the nonlinear Schrödinger equation, is a universal wave equation that describes a variety of systems, such as optical fields in nonlinear Kerr media~\cite{agrawal2000nonlinear}, weakly interacting Bose-Einstein condensates~\cite{kevrekidis2008emergent}, and gravity waves in deep inviscid fluids~\cite{craig1992nonlinear}. We focus on the system of the weakly interacting Bose gas in 3D, where 
$\psi(\boldsymbol{r},t)$ is interpreted as the classical field of the Bose gas, 
$g=4\pi\hbar^2 a/m$ is the strength of the interatomic interactions, $m$ is the atomic mass, and $a$ is the s-wave scattering length~\footnote{{It is worthwhile noting that, as in the recent experiments~\cite{Navon:2016,Dogra:2023}, our study remains in a regime $ka\ll 1$, where $k$ is the highest momentum considered. The scattering cross-section is thus momentum independent and $g=4\pi\hbar^2 a/m$.}}. 
The field is normalized to $\int |\psi(\boldsymbol{r},t)|^2\text{d}^3\boldsymbol{r}=N$, where $N$ is the (instantaneous) particle number. 

To study universal steady-state properties of the GP model far from equilibrium, one has to supplement forcing and dissipation mechanisms into Eq.~(\ref{eq:GP0}); here we do that by adding a potential term of the form $V(\boldsymbol{r},t)=V_\text{drive}(\boldsymbol{r},t)+V_\text{diss}(\boldsymbol{r})+V_\text{box}(\boldsymbol{r})$ (see~\cite{Supplementary}). The first term is a forcing at the length scale of the system size $L$, $V_\text{drive}(\boldsymbol{r},t)=\Us \sin(\omega t)z/L$.  The second term implements small-length-scale dissipation. This dissipation term is critical for realizing a steady state in a continuously forced system; we choose $V_\text{diss}$ to mimic the dissipation encountered in experiments, \emph{i.e.}\ evaporative losses when the atom energy exceeds an energy $\UD$~\cite{Navon:2019}. The last term, $V_\text{box}$, is a confining potential; for experimental relevance we pick a cylindrical box potential~\cite{Navon:2021,Dogra:2023} whose axis is oriented along the shaking direction $z$ [see the cartoon of Fig.~\ref{fignk}(a)]~\footnote{Note that even though the trap shape affects large-scale properties, the far-from-equilibrium state at smaller scales is largely insensitive to the details of the confining potential.}. Recent experimental studies have confirmed that this classical field approach is useful for describing a wave-turbulent scale-invariant steady state of a quantum degenerate Bose gas~\cite{Navon:2016,Navon:2019}.

The system is initialized in the ground state of Eq.~(\ref{eq:GP0}) including $V_\text{box}(\boldsymbol{r})$, which corresponds to a Bose-Einstein condensate with a nearly uniform density, except near the boundary of the box. For relevance, we use numbers typical of recent experiments~\cite{Dogra:2023}: a box radius $R=15\,\mum$ and length $L=50\,\mum$, an initial atom number $N(t=0)=2\times 10^5$, and $a/a_0$ between $25$ and $400$ (where $a_0$ is the Bohr radius). The natural energy scale of the system $\zeta\equiv gn(t=0)$ therefore varies between $\kB\times 1.2\,{\rm nK}$ and $\kB\times19\,{\rm nK}$, the corresponding timescale $\tau\equiv h/\zeta$, between $2.6\,\textrm{ms}$ and $42\,\textrm{ms}$, and the natural lengthscale $\xi\equiv\hbar/\sqrt{2m\zeta}$, between $0.6\,\mum$ and $2.3\,\mum$ (using the mass $m$ of $^{39}$K); $n=N/V$ is the average density, where $V=\pi R^2L$ is the box volume.  The gas is driven at a frequency $\omega/(2\pi)$ that matches the resonance of the lowest-lying Bogoliubov excitation~\cite{Garratt:2019}; for our set of parameters, $\omega/(2\pi)$ ranges from $5$ to $20\,\textrm{Hz}$.

\textit{Momentum distribution.} As shown in Fig.~\ref{fignk}(b), the injection of energy results in a cascade front propagating to higher momenta, until hitting the momentum $\kD$~\footnote{{For convenience, we refer to the wavenumber $\textbf{k}$ as momentum, even though formally the momentum is $\textbf{p}=\hbar \textbf{k}$.}} at a time $t\approx\tD$~\footnote{In End Matter~\cite{Supplementary}, we derive an analytic expression for $\tD$ in terms of the cascade properties.}.   
For $t\gtrsim\tD$, the system is in a steady state well described by a power law distribution for the mode occupation number: $N
_k \propto k^{-\gamma}$ with $\gamma\approx3.5$. Here, $N_k$ is the mode occupation number for the states of momentum $k$, normalized as $\sum_{\boldsymbol{k}} N_k=N$; in the continuous limit, it is related to the momentum distribution $n(k)$ as $N_k=((2\pi)^3/V)n(k)$.

We systematically studied the momentum-resolved cascade exponent $\gamma(k)\equiv-\textrm{d}\ln[n(k)]/\textrm{d}\ln{[k]}$ across interaction strengths (different $\xi$) and find that $\gamma(k\xi)$ is in excellent agreement with the $4$-wave WWT prediction $3+1/(3\ln[k/\kF])$, provided that $\kF$ is replaced by $k_0 = 1.64(2)k_\xi$ (see End Matter and Fig.~\ref{figGamma}). In other words, the isotropic $4$-wave cascade's effective injection scale is set by the GPE scale $1/\xi$ instead of the physical injeciton scale $\kF$.
However, as was already presumed experimentally~\cite{Navon:2016,Dogra:2023}, we find that a power law $n(k)\propto k^{-\gamma_0}$ with an effective exponent $\gamma_0=3.5$ accurately captures all the relevant details, since the variation of the logarithmic correction for $\gamma(k)$ is small over the relevant momentum range~\footnote{Note that~\cite{Zhu:2023a} speculated that the steeper effective exponent could be due to the logarithmic correction. However, this work assumes that there is no BEC, but the presence of the BEC is crucial for understanding the effective injection scale $k_0\propto k_\xi$, hence~\cite{Zhu:2023a} cannot make quantitative comparisons to the experiment~\cite{Navon:2016}.} (see also a discussion in ~\cite{Supplementary}). From now on, we fix $n(k)$ to this power law and define the amplitude of the power law $n_0\equiv N_kk^3 (k\xi)^{\gamma_0-3}$.

\begin{figure*}[t!]
\centerline{\includegraphics[width=1\textwidth]{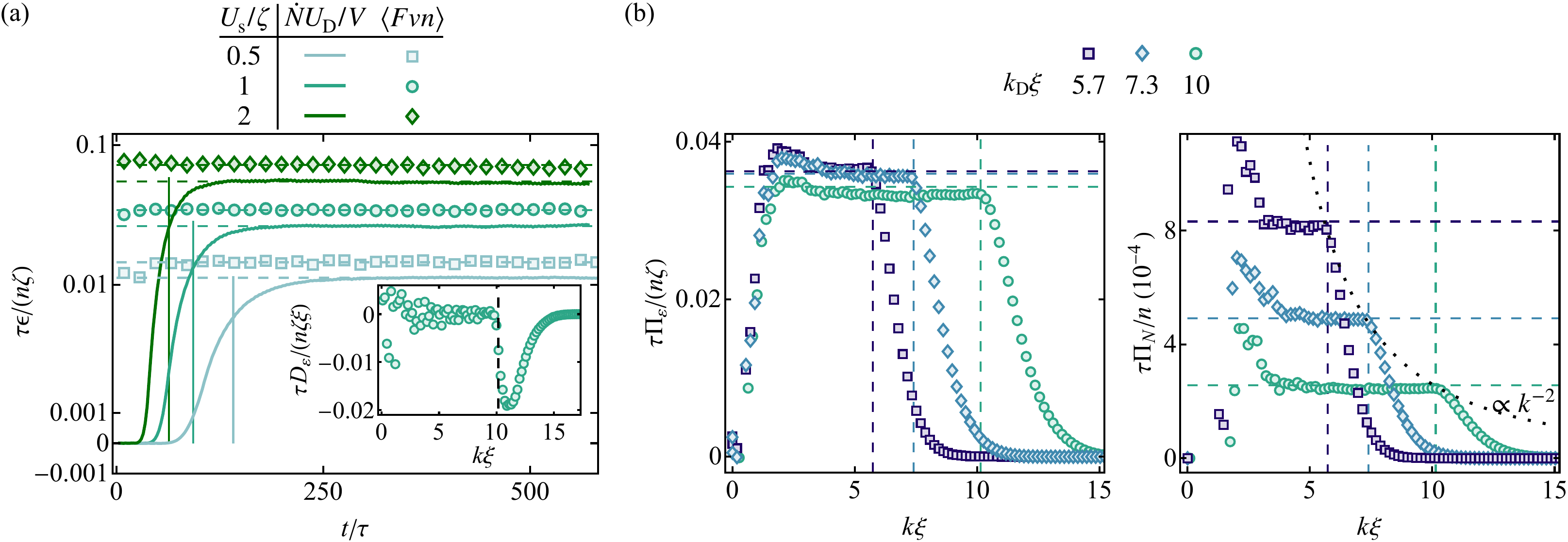}}
\caption{
Energetics of the direct turbulent cascade. (a) Energy input and dissipation rates. We show $\tau\epsilon/(n\zeta)$ where $\epsilon$ is either the energy injection rate calculated as $\Ein\equiv \langle Fvn\rangle$ (symbols) or the particle dissipation rate $\dot{N}$ multiplied by $\UD/V$ (solid lines). Both $\Ein/n$ and $\dot{N}\UD/N$ are constant at long times (dashed lines; see also~\cite{Supplementary}), but $\Ein/n$ is higher by a factor of $\approx1.3$. The vertical solid lines mark the onset time of dissipation $\tD$; for an analytical calculation of $\tD$, see the End Matter. Inset: the dissipation spectrum $D_\varepsilon$ for $\Us=\zeta$ and $a=100a_0$. The average dissipation momentum $\langle k_{\textrm{diss}}\rangle\approx1.15\kD$, predicting $\epsilon V/(\dot{N}\UD)\approx1.32$ (see text).
(b) Energy flux $\Pi_\varepsilon$ (left) and particle flux $\Pi_N$ (right) for different dissipation scales $\kD$ (vertical dashed lines). Both fluxes are scale independent. The dotted line ($\propto k^{-2}$) shows that $\Pi_N\propto \kD^{-2}$, while $\Pi_\varepsilon$ is (nearly) independent of $\kD$; horizontal dashed lines are $\Ein$ (resp. $\tau\dot{N}/N$) in the left (resp. right) panel (see text).
}
\label{figFlux}
\end{figure*}

\textit{Energetics of the turbulent cascade.} We next turn to the calculation of the energy-density input and dissipation rates. As energy is injected into the system only by the forcing $V_\text{drive}$, the energy input rate can be calculated as $\Ein\equiv\langle Fvn\rangle$ where $F\equiv-\hat{z}\cdot \nabla V_\text{drive}$, $\hat{z}$ is the unit vector along $z$, $v$ is the center-of-mass velocity of the gas and the averaging $\langle\cdot\rangle$ is performed over a drive period. The energy is dissipated at high momenta solely by particles with momenta $k>\kD$ leaving the trap. Hence, in previous experimental works~\cite{Navon:2019,Dogra:2023}, the energy dissipation rate was reasonably assumed to be $\dot{N}\UD$ where $\dot{N}$ is the particle loss rate. However, as shown in Fig.~\ref{figFlux}(a) for three different $\Us$, the energy-density injection rate $\Ein$ is consistently higher than $\dot{N}\UD/V$. 
The reason for $\Ein>\dot{N}\UD/V$ stems from the shape of the dissipation spectrum $D_\varepsilon(k)$ (see~\cite{Supplementary} for a formal definition): calculating the energy-density dissipation rate as $\dot{N}\UD/V$ assumes that the dissipation happens exactly at energy $\UD$. As shown in the inset of Fig.~\ref{figFlux}(a) for $\Us=\zeta$, $D_\varepsilon(k)$ has a sharp onset at $\kD$, but has a significant tail above $\kD$. The average dissipation momentum is $\langle k_{\textrm{diss}}\rangle\equiv\int k D_\varepsilon(k)\textrm{d}k/\int D_\varepsilon(k) \textrm{d}k \approx1.15\kD$~\footnote{A simple geometric picture is that particles hitting the box at non-normal incidence can remain trapped for some time even if their kinetic energy $\hbar^2 k^2/(2m)>\UD$.}, showing that the energy dissipated from the system per unit time is $\approx 1.32\dot{N}\UD$. This is in excellent agreement with $
\alpha\equiv\Ein/(\dot{N}V\UD)$~\footnote{Even though $\alpha\neq 1$ means that the energy-density flux is not $\dot{N} \UD/V$, the fact that $\alpha$ has only a very weak dependence on $a$ ($\propto a^{0.08(1)}$) and is independent of $\Us$ suggests that the determination of the energy flux via $\dot{N} \UD/V$, as done in~\cite{Navon:2019,Dogra:2023}, will only introduce a small, presumably system-specific, systematic error.} [see the End Matter for a systematic study of $\alpha$ versus interaction strength and $\Us$, showing that $\alpha\approx1.3$].

Having verified that the energy input rate (at low $k$) and dissipation rate (at high $k$) are equal, we turn to the direct calculation of the scale-resolved energy flux $\Pi_\varepsilon(k)$ and particle flux $\Pi_N(k)$ (see~\cite{Supplementary} for the formal definition of the fluxes). As shown in Fig.~\ref{figFlux}(b) in the steady state, $\Pi_\varepsilon$ is momentum independent in the \emph{inertial range}, a certain range of momenta where neither forcing nor dissipation takes place;
this flux transports the energy injected at low $k$ to higher $k$, up to $\kD$.
Importantly, $\Pi_\varepsilon$ in the inertial range is equal to $\Ein$.
As expected for an energy cascade, $\Pi_\varepsilon$ is also (nearly) independent of the dissipation scale $\kD$~\footnote{Note that the slow decrease of the plateau of $\Pi_\varepsilon$ is due to a progressive depletion of the condensate, which results in a (slow) decrease of $\Ein$ [corresponding horizontal dashed lines in the left panel of Fig.~\ref{figFlux}(b)].}. On the other hand, $\Pi_N$ is also momentum independent in steady state, but its plateau value decreases as $\propto \kD^{-2}$, as expected for particles with a quadratic dispersion relation [see dotted line in the lower panel of Fig.~\ref{figFlux}(b)]~\footnote{Note that it is only in the limit $\kD\rightarrow\infty$ that one recovers the expected limits $\Pi_\varepsilon\rightarrow \epsilon$ with $\Pi_N\rightarrow 0$ for the direct energy cascade~\cite{Dyachenko:1992}.}. 
We define $\epsilon$, the scale-invariant energy-density flux, as the plateau value of $\Pi_\varepsilon$; in the rest of the paper we compute $\epsilon$ as $\langle Fvn\rangle$.

\begin{figure*}[t!]
\centerline{\includegraphics[width=1\textwidth]{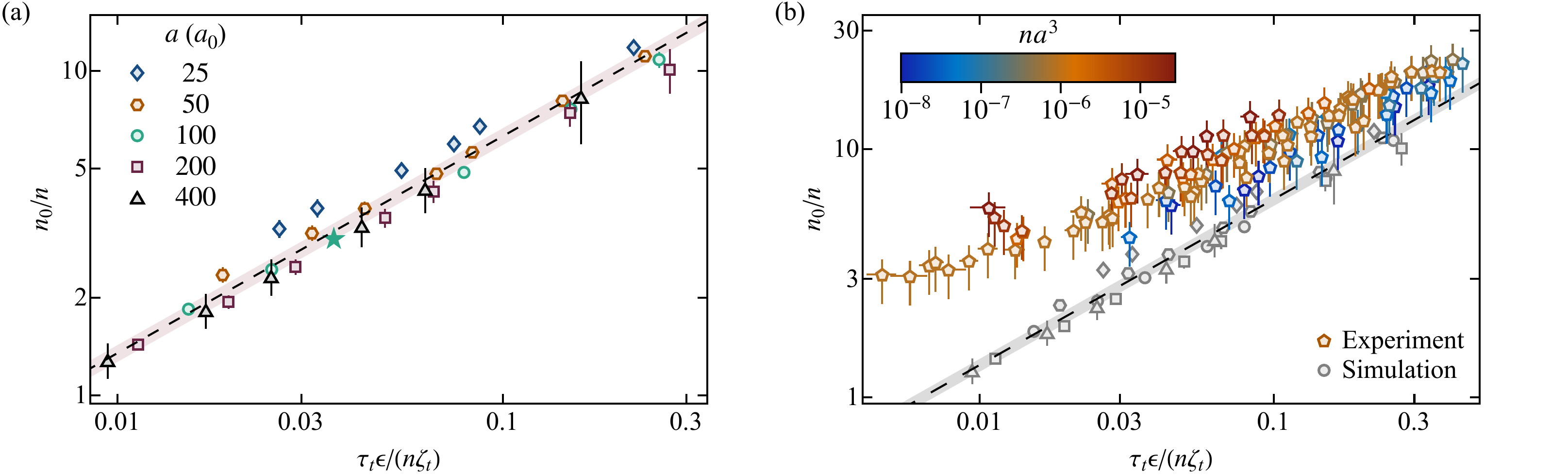}}
\caption{A universal equation of state of the GP model. (a) The numerical calculations for different $a$ collapse onto a universal curve when the state variables $\epsilon$ and $n_0$ are expressed in the instantaneous natural scales of the GP model ($n$, $\xi_t$, $\zeta_t$, $\tau_t$).
The dashed line is a power-law fit to the data that gives $n_0/n=29(2)(\tau_t\epsilon/[n\zeta_t])^{0.67(2)}$, and the pink band shows the fit uncertainty. (b) The comparison of our numerical EOS (gray symbols) with the experimental data from~\cite{Dogra:2023} (colored symbols). The color coding of the experimental data is based on the gas parameter $na^3$; note that the experimental fluxes are multiplied by $1.3$ compared to the data of~\cite{Dogra:2023} to account for $\alpha\neq1$.
The vertical error bars of the experimental data represent the uncertainties due to different $\gamma_0$ in simulations and experiments (see~\cite{Supplementary}).
}
\label{figEoS}
\end{figure*}

\textit{EOS for turbulence.} Finally, we investigate the relation between the two far-from-equilibrium state variables of the system, $\epsilon$ and $n_0$. Fig.~\ref{figEoS}(a) shows $n_0$ and $\epsilon$ for different $a$; we express both state variables using the instantaneous natural scales of the GP equation $\xi_t$, $\tau_t$, and $\zeta_t$ defined through the instantaneous density $n$~\footnote{Note that $\xi_t$ is not directly present in the axes of Fig.~\ref{figEoS}, but is used in the calculation of $n_0=N_kk^{-3}(k\xi_t)^{-0.5}$.}. Data for different values of $a$ fall on a (nearly) universal curve, demonstrating that the only relevant scales describing the turbulent cascade are the intrinsic scales of the universal GP model Eq.~(\ref{eq:GP0}) and all the dependence on the system and drive scales ($L$, $R$, $\Us$, $\omega$, $\kD$) drops out~\footnote{{The independence of our EOS with $\kD$ presumably will not hold for large enough $\kD$ (beyond what we studied), as it is reasonable to expect that as $k\rightarrow\infty$ (with $k<\kD$), WWT results should be recovered.}}.

The data of Fig.~\ref{figEoS}(a) show that the GP model contains a turbulent EOS that is a power law of the form $n_0/n=C(\tau_t\epsilon/[n\zeta_t])^b$, with $C=29(2)$ and $b=0.67(2)$, which is not described by known paradigms of turbulence. The two broad paradigms are wave (compressible-energy dominated) and vortex (incompressible-energy dominated) turbulence. Our $b$ is inconsistent with any wave-kinetic description, as a kinetic theory with $\ell$-wave interactions predicts an exponent $b=1/(\ell-1)\leq0.5$. 
Furthermore, the most plausible $4$-wave theory predicts that the relation between $n_0$ and $\epsilon$ would have to be independent of the total density $n$, which is not the case for the EOS constructed here~\footnote{For an extensive comparison of our EOS to the prediction of $4$-wave theory (as well as how the respective $N_k$ can be deduced), see Section~VII and Fig.~\ref{figLogVsGamma}}
~\footnote{The factor of $\xi^{0.5}$ on the $y$-axis is a finite-$\kD$ effect, but even in its absence our EOS would depend on $n$; within the GP model, only the WWT prediction of a power-law EOS with an exponent of $1/3$ is independent of $n$.}. 

To go further, we decompose in~\cite{Supplementary} the energy spectrum into compressible and incompressible parts~\cite{Nore:1997} and show that while the compressible part is typically larger than the incompressible one in the $k$ range used to extract $n_0$, the two are of the same magnitude; moreover, the incompressible part increases more rapidly with $\epsilon$ (similar observations were also recently reported in~\cite{Fischer:2024}). 
The observed EOS is therefore likely a result of an interplay between compressible (wave) and incompressible (vortex) excitations. As far as we are aware, there is no existing theory that describes this regime of `mixed' quantum turbulence and the shape of the EOS remains to be understood. We further note that the value of $b$ being the same as the exponent of the scaling of the energy spectrum for vortex (\emph{i.e.}\ hydrodynamic) turbulence $E_\text{K41} \propto \epsilon^{2/3}$~\cite{Frisch:1995} is likely coincidental, as this prediction is for the incompressible energy spectrum while our EOS is for the momentum distribution; besides, our computed incompressible energy spectra are also distinct from the K$41$ prediction~\cite{Supplementary}.
We note, however, that Kolmogorov turbulence can also be realized and measured in ultracold-atom experiments~\cite{Zhao:2025} (for $k\xi\ll 1$) but in that case the K$41$ spectrum originates from (coarse-grained) velocity field rather than the (atomic) momentum distribution.

\begin{figure}[t!]
\centerline{\includegraphics[width=1\columnwidth]{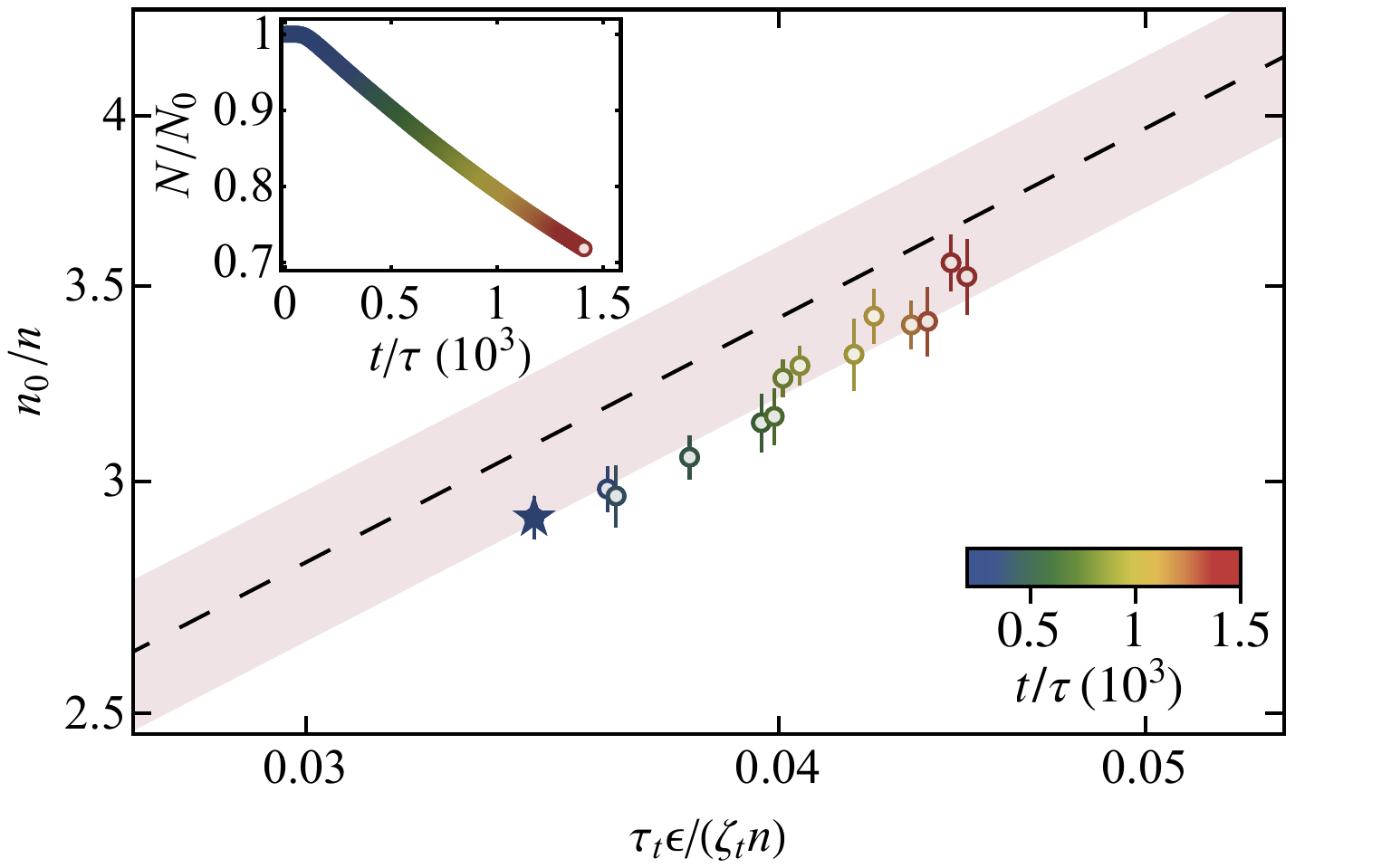}}
\caption{
Quasi-static process far from equilibrium. The instantaneous far-from-equilibrium state variables for different $t$ follow the EOS, in analogy with a  quasi-static process in equilibrium thermodynamics. The data correspond to $a=100a_0$
and the blue star here corresponds to the green star in Fig.~\ref{figEoS}(a). 
The inset shows the fraction of particles $N/N_0$ left in the system.
}
\label{figQuasistatic}
\end{figure}

In Fig.~\ref{figEoS}(b) we compare our calculations (grey symbols) with the experimental measurements~\cite{Dogra:2023} (colored symbols); see also~\cite{Supplementary} for details. The experimental data are above the numerical results. Note that the experimental data
do not represent a universal EOS on this plot: they collapse onto a single curve when both axes are rescaled with empirically determined powers of the gas parameter $na^3$~\cite{Dogra:2023}, a scaling that is inherently beyond the GP model. However, we note that in the limit $n\rightarrow \infty$ and $na^3\rightarrow 0$ the GP model is expected to be a good description of the experimental setting and Fig.~\ref{figEoS}(b) shows that the experimental data appear to approach our numerical results for lower $na^3$.

Finally, we note that the slow decrease of $N$ due to evaporative losses above $\kD$ implements a slow thermodynamic-like process. 
Indeed, as $N$ decreases, the state variable $\epsilon\propto N$ also decreases, resulting in a slow change of the far-from-equilibrium state (see also~\cite{Supplementary}). In Fig.~\ref{figQuasistatic}, we show that the simultaneous change of the state variables $\epsilon$ and $n_0$ follows the EOS: when rescaled to the instantaneous GP scales, the instantaneous $\epsilon$ and $n_0$ `slide up' on our universal EOS line over time, see blue to red shades~\footnote{As shown in~\cite{Supplementary}, $n_0$ and $\epsilon/n$ are roughly constant over time, resulting in the increase of $n_0/n$ and $\tau_t\epsilon/(n\zeta_t)$.  The points in Fig.~\ref{figEoS}(a) correspond to shaking times $\approx2\tD$.}. This is reminiscent of the concept of thermodynamic quasi-static processes, for which infinitesimal changes of external constraints take a system through a dense succession of equilibrium states~\cite{Callen:1991}. Remarkably, we find that this concept generalizes to states that are even locally far from equilibrium.

\textit{Conclusions.} We numerically investigated the properties of a turbulent cascade arising in the GP model when a box-trapped gas is periodically driven, and showed that it can be described by a universal EOS relating the turbulent state variables, the energy flux and the cascade amplitude.
The form of our EOS is inconsistent with any existing theory of turbulence, posing a new theoretical challenge. Furthermore, the comparison of our classical-field simulations to experimental measurements provides valuable benchmarks for testing the validity of classical-field theories in far-from-equilibrium scenarios. 
Namely, we find that the GP model provides valuable insight into the experimental results of~\cite{Dogra:2023}, even if it fails to fully capture the equation of state -- indeed, the observed $na^3$ scaling is incompatible with the universal GP framework (see also~\cite{Supplementary}).
This partial success is nevertheless striking given that there is no firm theoretical justification for its applicability in this turbulent regime. Understanding the foundations of this success, and developing refined approaches that incorporate quantum effects, remain important challenges for the field.
One should also investigate the fluctuations of $N_k$; comparing these fluctuations between experiments and classical field simulations could shed light on the role of quantum fluctuations in turbulent quantum fluids.

We thank M. Gazo, J. Etrych, C. Eigen, G. Krstulovic, M. Tsubota, G. Falkovich, and V. L'vov for discussions. This work was supported by the NSF (Grant Nos. PHY-1945324 and PHY-2110303), DARPA (Grant No. HR00112320038), the David and Lucile Packard Foundation, the Alfred P. Sloan Foundation, and JSPS KAKENHI (Grant No. JP23K13029). G.M. acknowledges support from Trinity College (Cambridge) and Yale University for its hospitality.

\clearpage
\newpage

\section{End Matter}
\label{EndMatter}

\begin{figure}[h!]
\centerline{\includegraphics[width=1\columnwidth]{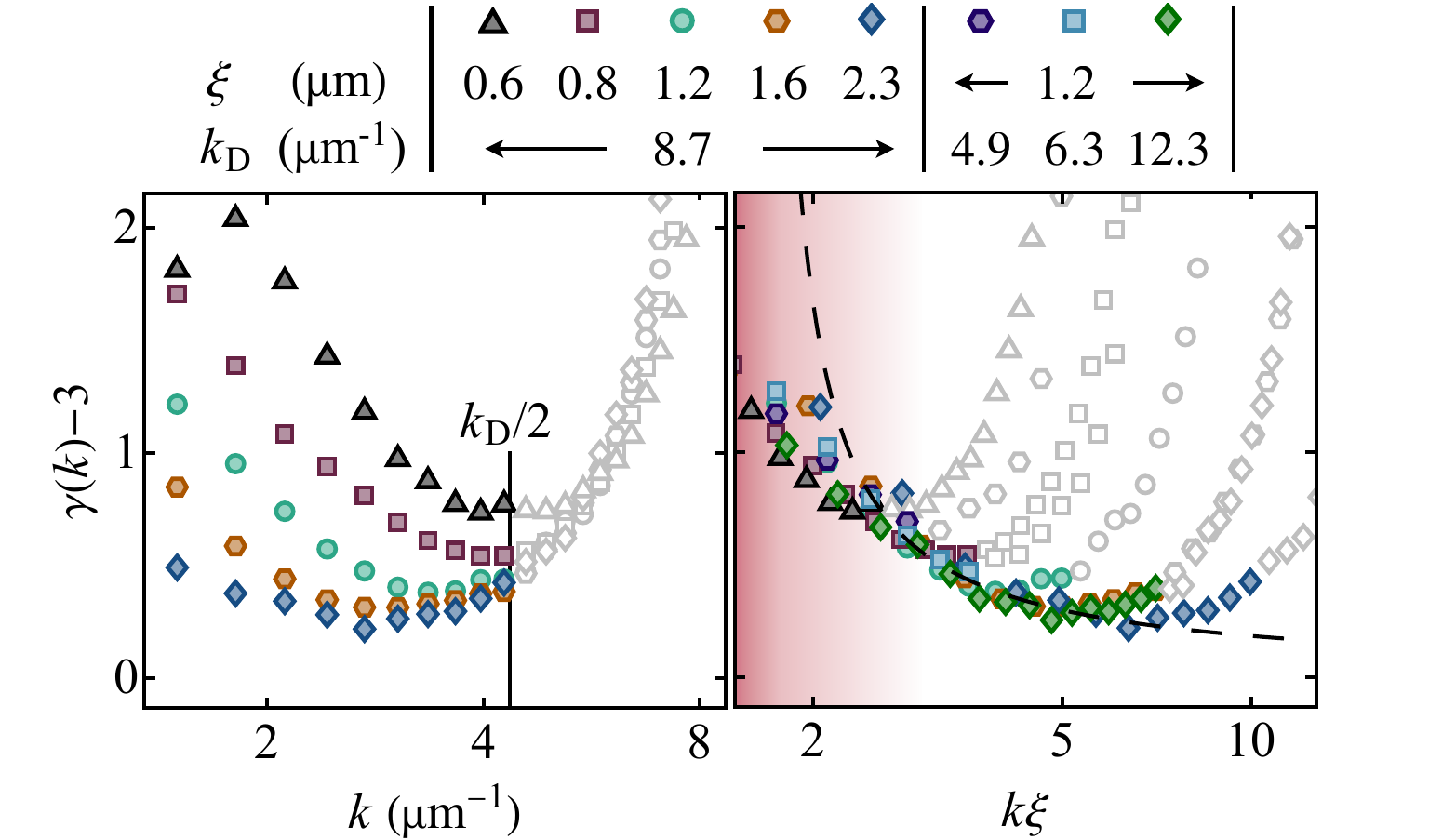}}
\caption{The cascade exponent $\gamma(k)$ for different system parameters. (left) $\gamma(k)$ for different interactions is not universal, and it bends up around $\kD/2$ (solid line). (right) The rescaled data for $\gamma$ versus $k\xi$ collapse in the range $2.5\lesssim k\xi\lesssim\kD\xi/2$, demonstrating that the effective injection scale $k_0$ of the isotropic $4$-wave cascade is $\propto k_\xi\equiv1/\xi$. The dashed line shows the theoretical prediction $\gamma(k)-3=1/[3\ln(k/k_0)]$~\cite{Nazarenko:2011,Zhu:2023a} with $k_0=1.64k_\xi$, and the red shading indicates the region of momentum space where the weak interaction approximation is not valid (see text). 
}
\label{figGamma}
\end{figure}

\emph{The cascade exponent $\gamma$.} In any real system, the momentum range over which universal turbulent properties (\emph{i.e.} injection and dissipation independent, boundary condition independent, \emph{etc.}) may exist is finite, and the observations might depend on the separation between the injection and dissipation scales. To test the universality of our steady-state $n(k)$, we calculate the apparent cascade exponent
$\gamma(k)\equiv-\textrm{d}\ln[n(k)]/\textrm{d}\ln[k]$ for different $\xi$ and $\kD$. As shown in Fig.~\ref{figGamma}, when $\gamma$ is plotted versus $k$ for various $\xi$ (but same $\kD$), it decreases for $k\lesssim\kD/2$ and depends on $\xi$; for $k\gtrsim\kD/2$, finite-$\kD$ effects become important and all $\gamma(k)$ increase together. 
However, when plotted against $k\xi$ -- the only dimensionless parameter for the infinite-size system -- the data collapse onto a single curve in the intermediate regime $2.5\lesssim k\xi\lesssim\kD\xi/2$~\footnote{One might expect $\gamma(k)$ to be universal in the range $k_\mathrm{F}\ll k\ll \kD$. 
We instead observe universal behavior for $k\xi\gtrsim2.5$, where $n(k)$ is isotropic~\cite{Galka:2022,Sano:2022} and $\gamma(k)$ is well defined. For $k\xi\lesssim2.5$, our numerical resolution of $\gamma(k)$ is insufficient to make a definitive claim.}. 

As the momentum $k_\xi\equiv1/\xi$ marks the typical momentum scale associated with interactions,
we have that for $k\gg k_\xi$, our system is weakly nonlinear and can be described by the theory of WWT~\cite{Zakharov:1992,Nazarenko:2011}. In this weakly interacting limit (not to be confused with the criterion $na^3\ll 1$), Eq.~(\ref{eq:GP0}) reduces to so-called $4$-wave (particle-number conserving) interactions~\cite{Nazarenko:2011}, and WWT in this case predicts that the momentum distribution of the direct energy cascade has the asymptotic isotropic form $n(k)\propto k^{-3}\ln^{-1/3}(k/k_0)$ where $k_0$ is the energy injection scale~\cite{Zhu:2023a}~\footnote{At lower momenta ($k\lesssim k_\xi$) the particles interact more strongly, and the $4$-wave WWT description is not expected to work. In the presence of a condensate, the proper quasiparticles to consider for $k\xi\lesssim1$ are the Bogoliubov phonons, for which the appropriate WWT description is a wave-kinetic equation with 3-wave interactions. In our simulations, a description in terms of an equilibrium-like condensate is meaningless since for our strong drives, the Bogoliubov approximation is invalid.}.

The collapse in Fig.~\ref{figGamma} shows that, despite the fact that the energy is physically injected at the scale $\kF\approx\pi/L$, the effective injection scale for the isotropic cascade is actually $k_0\propto k_\xi$ ($\gg \kF$)~\footnote{Interestingly, a recent experiment has observed that the momentum distribution of $2$D shaken Bose gases also becomes isotropic at $k\approx k_\xi$~\cite{Galka:2022}.}.
To test the asymptotic form $n(k)\propto k^{-3}\ln^{-1/3}(k/k_0)$, we fit $\gamma(k)$ with $3+1/(3\ln[Ak\xi])$ in the regime where the numerical data overlap. We find that for $A=0.61(1)$, the fit captures the data well. Equivalently, this fit predicts that the injection scale for the isotropic $4$-wave cascade is $k_0=1.64(2)k_\xi$ (as long as $\kF\ll k_\xi$~\footnote{Experimentally, $\kD$ is bounded, so the separation between the injection scale $k_0\propto k_\xi$ and $\kD$ -- and hence the region where the $4$-wave WWT theory is applicable -- is reduced for stronger interactions.}). The convergence towards a robust steady state, \emph{i.e.}\ that is independent of $\xi$, $\omega$, $\kD$, and $\Us$, and that matches a universal expectation that is independent of the injection and dissipation mechanisms as well as the boundary conditions strongly suggests that we have access here to intrinsic properties of the universal GP model Eq.~(\ref{eq:GP0}).

\begin{figure}[t!]
\centerline{\includegraphics[width=1\columnwidth]{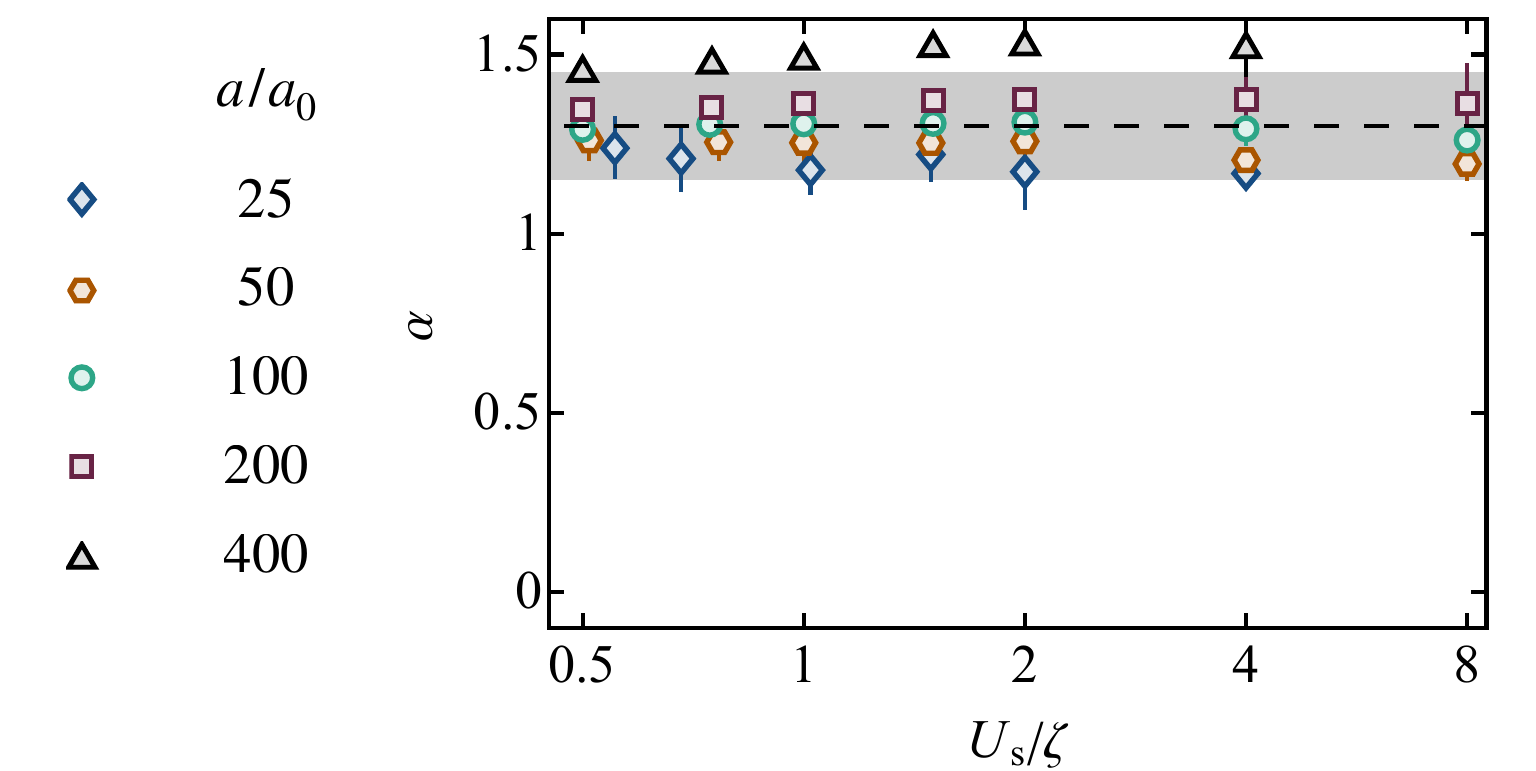}}
\caption{The ratio $\alpha$ of $\Ein$ and ${\dot{N}\UD/V}$ for different drive strengths $\Us/\zeta$ and scattering lengths $a/a_0$.
}
\label{figAlpha}
\end{figure}

\emph{Reconciling energy injection and dissipation rates.} Here we study the ratio between energy injection rate $\Ein$ and the \emph{apparent} energy dissipation rate $\dot{N}\UD/V$ systematically by showing their ratio $\alpha=V\Ein/(\dot{N}\UD)$ for different $\Us$ and $a$ in Fig.~\ref{figAlpha}. We find that the ratio is $\alpha \approx 1.3$, independent of $\Us$ and has a weak dependence on $a$.

\begin{figure*}[t!]
\centerline{\includegraphics[width=1\textwidth]{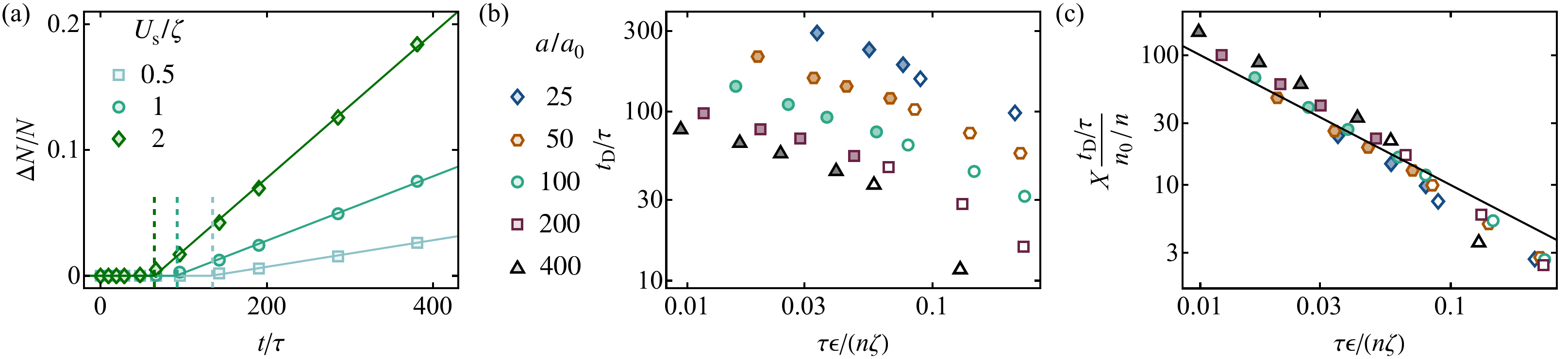}}
\caption{
The onset time for dissipation. (a) Number of particles lost from the system $\Delta N$ for different drive strengths $\Us/\zeta$; here $a=100a_0$. The lines are piece-wise linear fits, indicating that dissipation begins after a cascade build-up time $\tD$ (vertical colored dashed lines). (b,c) Onset time for dissipation $\tD$ as a function of $\epsilon$. (b) The relation is monotonic for a given interaction strength, but the data for $\tD$ at different interaction strengths are not universal. (c) The onset times can be analytically calculated assuming that $\epsilon$ is constant over time [for the explicit expression of $X$, see Eq.~(\ref{eq:tD})]. The solid black line is the theoretical expectation $n\zeta/(\tau\epsilon)$. The closed (resp. open) symbols correspond to driving with $\Us\leq1.5\zeta$ (resp. $\Us>1.5\zeta$). For $\Us\leq1.5\zeta$, the energy input rate is constant in time to within $10\%$.}
\label{figtD}
\end{figure*}

\emph{The onset time for dissipation.}
Here we calculate the time $\tD$ when the system starts to dissipate energy. As shown in Fig.~\ref{figFrontSharpness}, the cascade front is not very sharp, so the onset of dissipation is not very sharp either [see Fig.~\ref{figFlux}(a)]. However, looking at the fractional particle loss $\Delta N/N$ instead, the onset is more clearly identifiable; to define $\tD$, we fit $\Delta N/N$ with a piece-wise linear function and identify $\tD$ as the singular point of the piece-wise function [solid lines in Fig.~\ref{figtD}(a)].
In Fig.~\ref{figtD}(b), we show $\tD$ versus $\epsilon$ for different $a$; $\tD$ monotonically decreases with $\epsilon$ for a given interaction strength but the relation is not single valued for different $a$.

Remarkably, $\tD$ can be calculated analytically under the assumption that the energy input rate is constant and that the onset of dissipation is sharp at $k=\kD$. As the momentum distribution has the form $n(k)=V/(2\pi)^3n_0 k^{-3} (k\xi)^{-\gamma_0+3}$, the total energy of the system in the steady state is 
$$\int_0^{\kD} 4\pi k^2 n(k) \frac{\hbar^2k^2}{2m}\text{d}k=n_0\frac{\UD V (\kD\xi)^{3-\gamma_0}}{2\pi^2(5-\gamma_0)}.$$
Equating this to the energy injected into the system up until $\tD$ yields 
\begin{equation}
\label{eq:tD}
\frac{\tD}{\tau}=\frac{\UD (\kD\xi)^{3-\gamma_0}}{2\pi^2 (5-\gamma_0)\zeta}\frac{n_0/n}{\tau\epsilon/n\zeta}\equiv X\frac{n_0/n}{\tau\epsilon/n\zeta}.
\end{equation}
In Fig.~\ref{figtD}(c) we show that $\tD$ is in excellent agreement with this calculation for weak drives ($\Us\leq 1.5\zeta$), while for stronger drives the constant-$\epsilon$ assumption fails and $\tD$ is shorter [open symbols in Fig.~\ref{figtD}(c)]. 
Note that $\tD\propto \UD \kD^{3-\gamma_0}\propto \kD^{5-\gamma_0}$, is positive for our $\gamma_0=3.5$, meaning that $\tD\rightarrow \infty$ as $\kD\rightarrow\infty$. In the WWT language, this is referred to as an infinite capacity cascade~\cite{Nazarenko:2011}. This behavior is in stark contrast with the case of the K41 spectrum, where $\tD$ is finite as $\kD\rightarrow\infty$ (more specifically, $\tD\sim\tD^{(0)}- A\kD^{-2/3}$ where $A$ is a dimensionful constant).

Despite the weak nonlocality discussed in section~III, the dissipation onset time $\tD$ matches the analytical prediction. 
In Fig.~\ref{figFlux}(a), we see that the curves for $\dot{N}\UD/N$ are slightly rounded before reaching their steady-state value but $\tD$ is an approximate point of symmetry where the (unaccounted) energy dissipation before $\tD$ matches the not-yet-fully-saturated dissipation after $\tD$.

\newpage

\setcounter{figure}{0} 
\setcounter{equation}{0} 

\renewcommand\theequation{S\arabic{equation}} 
\renewcommand\thefigure{S\arabic{figure}} 
\onecolumngrid

\section{Supplemental Material}
\label{SI}


\subsection{\textsc{I. Simulation details}}
Our numerical simulations solve the time-dependent GP equation [Eq.~(\ref{eq:GP0})] using a pseudo-spectral method with the fourth-order Runge-Kutta time evolution. We add potential terms to the GP equation to implement the box trapping and to provide energy injection (which conserves particle number) and dissipation (which does not):
\begin{equation}
\label{eq:GP}
    i\hbar\frac{\partial \psi}{\partial t}=\left(-\frac{\hbar^2}{2m}\nabla^2+g|\psi|^2+V_\mathrm{box}(\boldsymbol{r},t)+V_\mathrm{drive}(\boldsymbol{r},t)+V_\mathrm{diss}(\boldsymbol{r},t)\right)\psi.
\end{equation}
These potentials are selected for their experimental relevance~\cite{Navon:2019,Dogra:2023}. The cylindrical box potential, whose depth is $\UD$ and symmetry axis is along $z$, is:  
$$V_\mathrm{box}(\boldsymbol{r},t)=
\begin{cases}
    0, & \text{if } |z|<L/2 \text{ and } x^2+y^2<R^2\\
    \UD,              & \text{otherwise,}
\end{cases}$$
$V_\mathrm{drive}$ is the time-periodic potential gradient:
$$V_\mathrm{drive}(\boldsymbol{r},t)=\frac{\Us}{L}\,\textrm{sin}(\omega t)z,$$
and $V_\mathrm{diss}$ is an absorbing (imaginary) potential on the edges of the simulation grid to mimic the loss of particles that have enough energy to leave the box:
$$V_{\mathrm{diss}}(\boldsymbol{r},t)=-\textrm{Max}\left[\textrm{cosh} ^{-2}
\left( \frac{1}{w}[1-|x|/D_x]\right),
\textrm{cosh} ^{-2}
\left( \frac{1}{w}[1-|y|/D_y]\right),
\textrm{cosh} ^{-2}
\left( \frac{1}{w}[1-|z|/D_z]\right)
\right]iA.$$ 
Here, the edges of the simulation grid are located at $\pm D_\sigma$ along direction $\sigma$, $A$ and $w$ are the strength and the characteristic width of the absorbing boundary. We benchmarked our simulations for different choices of $A$ and $w$ and have verified their choice within an appropriate range does not affect our results. Note that we do not include the gravitational potential; this is justified as in experiments in 3D optical boxes, gravity is well compensated (\emph{e.g.} by magnetic levitation)~\cite{Navon:2021}.

Our simulations are typically performed on a grid of size $128\times128\times256$ (for the data at the highest $\kD$ in Fig.~\ref{figGamma}, we use a $256\times256\times512$ grid to ensure that the highest accessible momenta in the simulations are $>\kD$). Our temporal resolution is $\approx0.1\hbar/\UD$.

\subsection{\textsc{II. Calculating momentum-resolved fluxes}}
The equation of motion for the energy-density spectrum, $\varepsilon(k,t)= 1/n\int_{|\boldsymbol{k^\prime}|=k}n({\boldsymbol{k}^\prime})\hbar^2|\boldsymbol{k}^{\prime}|^2/(2m)\text{d}\boldsymbol{k^\prime}$ (where the time dependence of $n({\boldsymbol{k}^\prime})$ is omitted for convenience) is obtained directly from the GP equation by taking the time derivative of $\varepsilon(k,t)$:
\begin{equation}
\label{eq:energyContinuity}
    \frac{\partial}{\partial t}\varepsilon(k,t)+\frac{\partial}{\partial k}\Pi_\varepsilon(k,t)=F_\varepsilon(k,t)+D_\varepsilon(k,t).
\end{equation}
The energy-density flux $\Pi_\varepsilon$ is 
\begin{equation}
\Pi_\varepsilon(k,t)=\frac{1}{i\hbar}\int_k^\infty \text{d}k_1\int_{|\boldsymbol{k^\prime}|=k_1}\frac{\hbar^2|\boldsymbol{k}^{\prime}|^2}{2m}(\Tilde{\psi}^*(\boldsymbol{k}^{\prime})\mathbb{FT}[g|\psi|^2\psi](\boldsymbol{k}^{\prime})-\text{c.c.})\text{d}\boldsymbol{k^\prime},
\label{eqPiEpsilon}
\end{equation}
where $\Tilde{\psi}(\boldsymbol{k})$ is the Fourier transform of $\psi(\boldsymbol{r})$ (the time dependence of $\tilde{\psi}$ and $\psi$ are omitted for convenience), $\mathbb{FT}$ denotes the Fourier transform, and $\text{c.c.}$ denotes the complex conjugate. The terms $F_\varepsilon$ and $D_\varepsilon$ are the (system-dependent) forcing and dissipation terms respectively:
$$F_\varepsilon(k,t)=\frac{1}{i\hbar}\int_{|\boldsymbol{k^\prime}|=k}\frac{\hbar^2|\boldsymbol{k}^{\prime}|^2}{2m}\left(\Tilde{\psi}^*(\boldsymbol{k}^{\prime})\mathbb{FT}\left[\frac{\Us}{L}z\sin(\omega t)\psi\right](\boldsymbol{k}^{\prime})-\text{c.c.}\right)\text{d}\boldsymbol{k^\prime},$$
and
$$D_\varepsilon(k,t)=\frac{1}{i\hbar}\int_{|\boldsymbol{k^\prime}|=k}\frac{\hbar^2|\boldsymbol{k}^{\prime}|^2}{2m}\left(\Tilde{\psi}^*(\boldsymbol{k}^{\prime})\mathbb{FT}[V_{\mathrm{box}}\psi+V_{\mathrm{diss}}\psi](\boldsymbol{k}^{\prime})-\text{c.c.}\right)\text{d}\boldsymbol{k^\prime}.$$
The analog of Eq.~(\ref{eq:energyContinuity}) for the particle density spectrum $1/n\int_{|\boldsymbol{k^\prime}|=k}n({\boldsymbol{k}^\prime})\text{d}\boldsymbol{k^\prime}$ is derived along similar lines, and the corresponding particle-density flux $\Pi_N$ is
$$\Pi_N(k,t)=\frac{1}{i\hbar}\int_k^\infty \text{d}k_1\int_{|\boldsymbol{k^\prime}|=k_1}(\Tilde{\psi}^*(\boldsymbol{k^\prime})\mathbb{FT}[g|\psi|^2\psi](\boldsymbol{k}^{\prime})-\text{c.c.})\text{d}\boldsymbol{k^\prime}.$$
The results presented in Fig.~\ref{figFlux}(a)-(b) are obtained by direct calculations of the above formulae.

\subsection{\textsc{III. Sharpness of the cascade front}}
\label{section:frontSharpness}
\begin{figure*}[t!]
\centerline{\includegraphics[width=\textwidth]{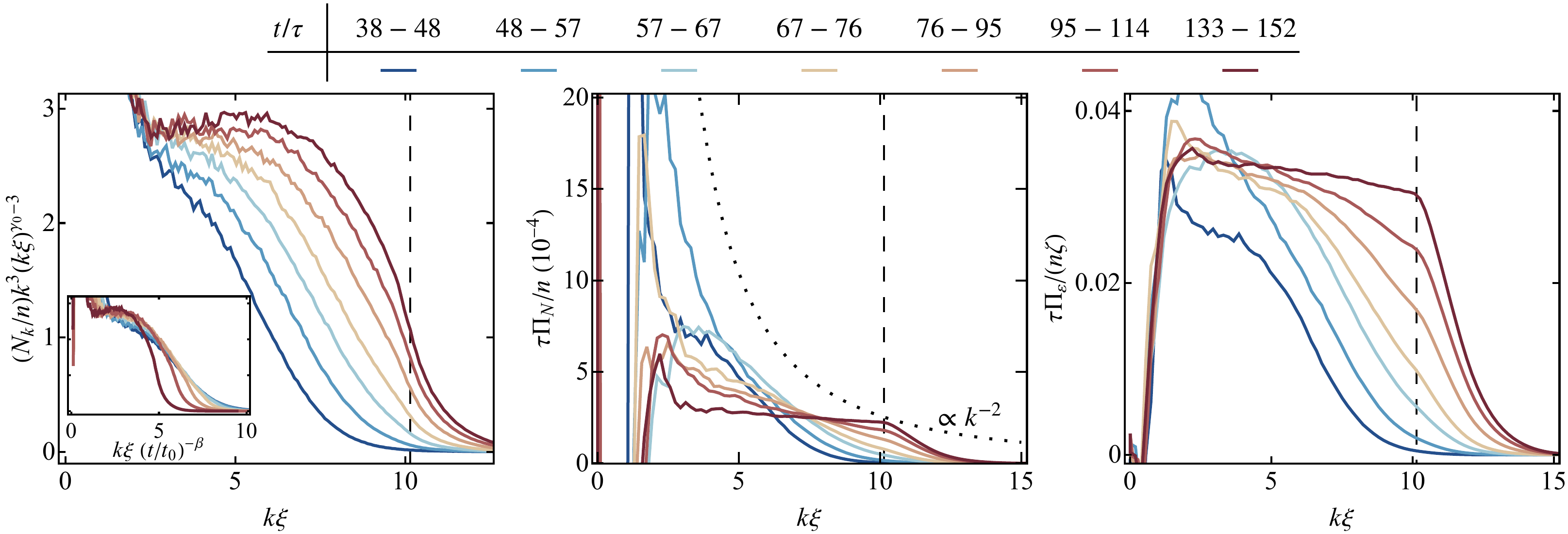}}
\caption{
Sharpness of the cascade front. 
The compensated spectrum $\propto N_k k^{\gamma_0}$ (left), the particle flux $\Pi_N(k)$ (middle), and the energy flux $\Pi_\varepsilon(k)$ (right) for various times. During the cascade buildup, neither the spectrum nor the fluxes have sharp fronts. The dissipation scale $\kD$ acts as a sharp cutoff in momentum space and results in a sharp drop-off of the fluxes above $\kD$. The inset on the left panel shows the compensated spectrum with the $x$-axis rescaled by $(t/t_0)^{-\beta}$ showing that the cascade propagation is only approximately dynamically scalable; here $\beta=2/3$ is calculated from $\gamma_0=3.5$, and $t_0=0.5\,\textrm{s}$ is an arbitrary reference time. The dotted line in the middle plot is a reference $\propto k^{-2}$, same as in Fig.~\ref{figFlux}(b).}
\label{figFrontSharpness}
\end{figure*}

In WWT theory, the interactions between waves are usually assumed to be local in momentum space, \emph{i.e.}\ interactions are significant only between waves of nearby momenta, resulting in a sharp cascade front propagating in momentum space. This locality arises from the conservation of energy and momentum: the most likely interactions satisfying those conservation laws are the ones that result in a small change of momenta. The effects of the conservation laws depend on the dimensionality and are more pronounced in lower dimensions. It turns out that WWT predicts that the direct energy cascade in the 3D GP model is not strictly local due to a weak (logarithmic) divergence at low $k$~\cite{Dyachenko:1992,Falkovich:1992,Zhu:2023a}. In Fig.~\ref{figFrontSharpness} we show that both the compensated spectrum $(N_k/n)k^3 (k\xi)^{\gamma_0-3}$ and the energy flux $\Pi_\varepsilon$ do not have very sharp edges before hitting $\kD$. Because of this, the energy dissipation rate $\dot{N}\UD$ grows smoothly from $0$ to its final value rather than sharply jumping at $\tD$ (see Fig.~\ref{figFlux}(a)). Consequently, the dynamics of the fluxes is more complex than the simple picture surmised in Fig.4 of~\cite{Navon:2019}. However, once the dissipation scale $\kD$ is reached, it enforces a sharp cutoff to the flux, resulting in the textbook behavior versus $\kD$ shown in Fig.~\ref{figFlux}(c).

\begin{figure*}[b!]
\centerline{\includegraphics[width=\textwidth]{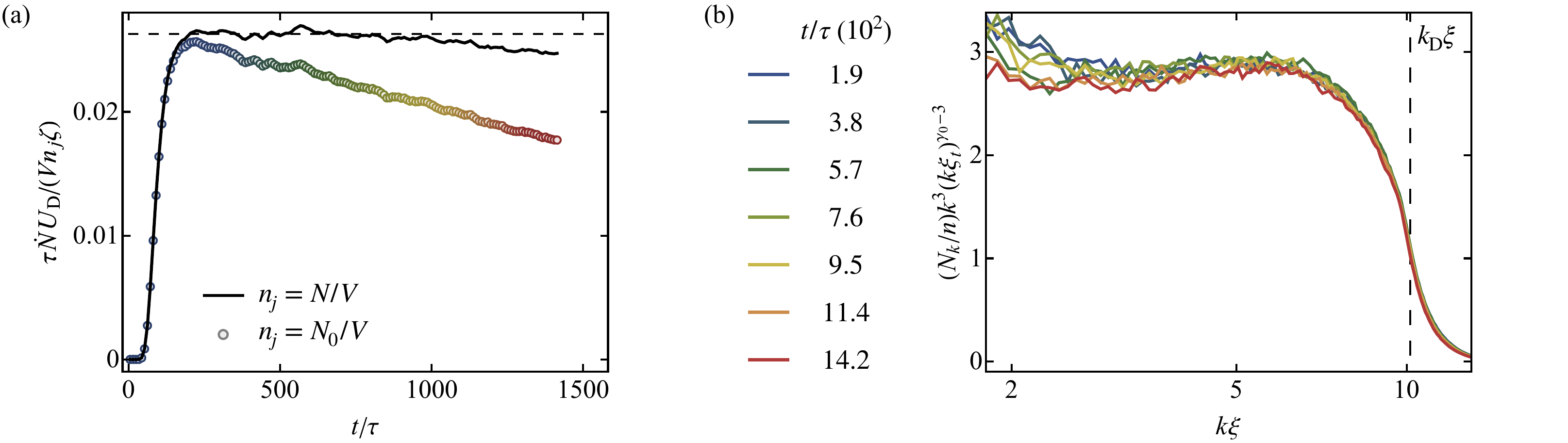}}
\caption{
Time dependence of state variables.
(a) The particle-density dissipation rate $\dot{N}/V$ multiplied by $\UD$ and normalized to $n_j$. As a black solid line (resp. colored symbols), we show $\dot{N}\UD/V$ normalized to $n_j=N/V$ (resp. $n_j=N_0/V$), the instantaneous density (resp. initial density). After the steady state is fully established, the former is constant while the latter decreases as atoms leave the trap.
(b) The compensated spectrum $(VN_k/N_0)(k\xi_t)^{\gamma_0-3}$ is constant over time, even though the energy flux has significantly decreased.
The color coding is the same in the two panels.
}
\label{figTimeDependence}
\end{figure*}

\subsection{\textsc{IV. Time dependence of the state variables}}

For long drive times, the atom loss becomes significant, resulting in a depletion of the low-$k$ population and thus a decrease in the total energy injection rate into the system [$\propto N$, Fig.~\ref{figTimeDependence}(a)]. On the other hand, the compensated spectrum $N_kk^3(k\xi_t)^{\gamma_0-3}$ is constant (Fig.~\ref{figTimeDependence}(b)). This can be understood from the shape of the EOS as follows: $\epsilon$ decreases as $1/N$, but $\tau_t/(n\zeta_t)$ increases as $N^3$, resulting in $n_0/n$ increasing as $N^{4/3}$ (assuming that the EOS is $n_0\propto \epsilon^{2/3}$). This means that $(N_k/n)k^3(k\xi_t)^{\gamma_0-3}$ increases as $N^{13/12}$ for $\gamma_0=3.5$, and therefore the compensated spectrum $N_kk^3(k\xi_t)^{\gamma_0-3}$ increases as $N^{1/12}$, too weak to be observed.

\subsection{\textsc{V. Comparison of our numerical EOS with the prediction of WWT theory}} 

\begin{figure*}[b!]
\centerline{\includegraphics[width=\textwidth]{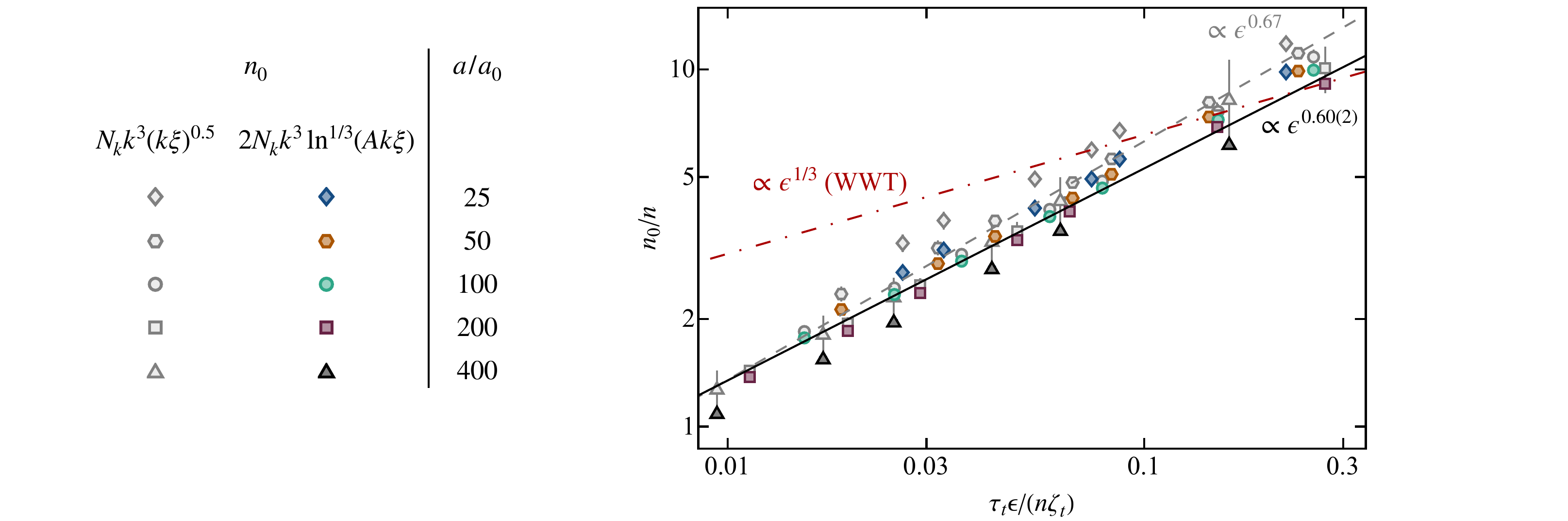}}
\caption{
Comparison of our numerical EOS with the predictions of WWT theory. For the gray symbols (same as in Fig.~\ref{figEoS}) the cascade amplitude $n_0$ is defined as $n_0=N_kk^{3}(k\xi)^{0.5}$ while for colored symbols as $n_0=2N_kk^3\ln^{1/3}(Ak\xi)$ with $A=0.61$. The black solid (resp. gray dashed) lines are power-law fits to the colored (resp. gray) symbols, giving a power law exponent $0.60(2)$ [resp. $0.67(2)$]. The red dash-dotted line shows the prediction of WWT theory for the $4$-wave direct cascade without any adjustable parameters.
}
\label{figLogVsGamma}
\end{figure*}

In the main text, we define the cascade amplitude as $n_0=N_k k^{3} (k\xi)^{\gamma_0-3}$; however, as shown in Fig.~\ref{figGamma}, $N_k$ is not exactly a power law and therefore the cascade amplitude should be more rigorously defined as $n_0^{\textrm{ln}}\propto N_kk^{3} \ln(k/k_0)^{1/3}$. Using the results of Fig.~\ref{fignk}(c), we define $n_0^{\textrm{ln}}=2N_kk^{3} \ln(k/k_0)^{1/3}$ with $k_0=1.64k_\xi$ (see Fig.~\ref{fignk}). In Fig.~\ref{figLogVsGamma} we show $n_0$ (as gray symbols) and $n_0^{\textrm{ln}}$ (as colored symbols), together with power-law fits to both. The two amplitudes are nearly identical, and the power-law fits give EOS exponents of $0.67(2)$ and $0.60(2)$ for $n_0$ and $n_0^{\textrm{ln}}$, so that the conclusions in the main text are unaffected (note that the factor of $2$ in the definition of $n_0^\textrm{ln}$ was chosen so that the amplitudes defined from the power law and with the $\ln$ correction are numerically close for our range of parameters; however, their respective dependence on $\epsilon$ is independent of that choice).

Additionally, we show in Fig.~\ref{figLogVsGamma} the $n_0^{\textrm{ln}}$ corresponding to the analytical prediction of the $4$-wave WWT cascade $N_k/n=(2\pi)^{8/3} C_\textrm{d}(\tau\epsilon/n\zeta)^{1/3}k^{-3}\ln[k/k_0]^{1/3}$ where $C_\textrm{d}\approx0.0526$~\cite{Zhu:2023a} and $k_0=1.64k_\xi$ (dash-dotted red line). The WWT predictions are similar to our observations, but the EOS exponent is clearly distinct.

Note that the results of Fig.~\ref{figLogVsGamma} enable the comparison of the predictions for $N_k$ as well. In the case of WWT, $N_k=C_{\rm d}/(2\pi)^3 n (\tau_t\epsilon/[2\pi n\zeta_t])^{1/3} k^{-3} \ln^{-1/3}{[A k \xi]}$, where $A=0.61$ (from Fig. 1), $C_{\rm d}=0.0526$~\cite{Zhu:2023a}, and the factors $2\pi$ come from different definitions used in our paper and Ref.~\cite{Zhu:2023a}. In the case of our simulations, $N_k=C(\tau_t\epsilon/[n\zeta_t])^b n k^{-3.5}$ with $C=29$ and $b=0.67$ in the inertial range.

\subsection{\textsc{VI. Comparison of our numerical EOS with the experimental measurements}}

Here we give additional details for the comparison of our numerical results to the experimental data from~\cite{Dogra:2023}. As the cascade exponent $\gamma_0$ used to define $n_0$ is different in the two cases ($\gamma_0=3.5$ for our simulations and $\gamma_0=3.2$ in~\cite{Dogra:2023}), the comparison of $n_0$ depends (weakly) on $k$. To convert $n_0$ reported in~\cite{Dogra:2023} to our definition with $\gamma_0=3.5$, we use $k$ in the middle of the experimental cascade range and use $k$ at the edges of the range to assess the uncertainty of the experimental data due to this correction; we use this uncertainty as an error bar for experimental points. We also multiply the experimentally measured fluxes by $\alpha=1.3$ to account for the difference between $\epsilon$ and $\dot{N}\UD/V$.

\subsection{\textsc{VII. Compressible and incompressible parts of the energy spectra}}
\begin{figure*}[h!]
\centerline{\includegraphics[width=\textwidth]{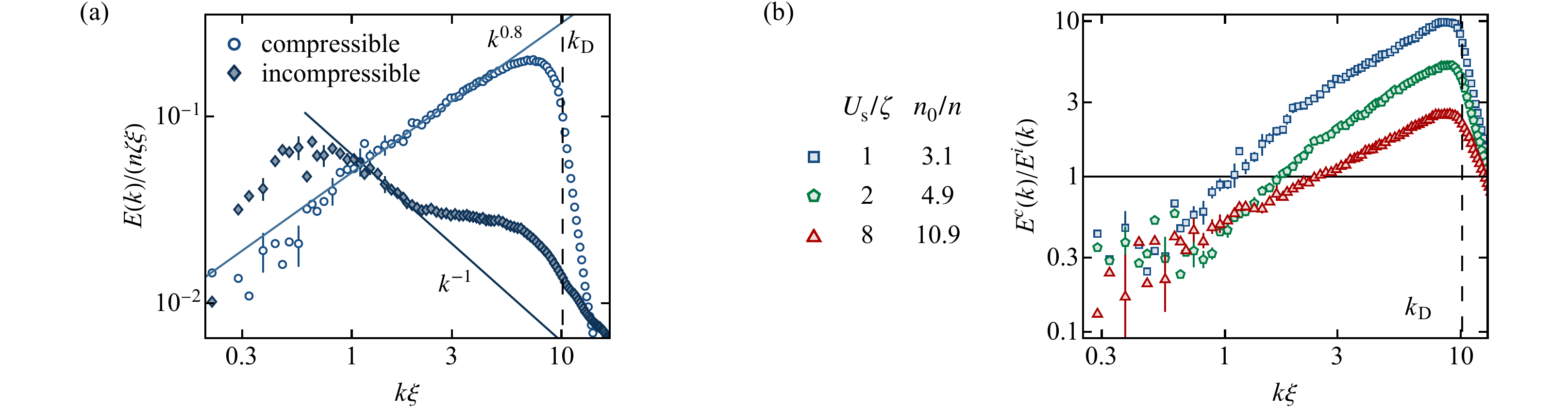}}
\caption{
Compressible $E^c(k)$ and incompressible $E^i(k)$ energy spectra for $a=100a_0$. (a) $E^c(k)$ and $E^i(k)$ for $\Us/\zeta=1$. The solid lines are power law guides to the eye. (b) The ratio $E^c(k)/E^i(k)$ becomes smaller for increasing $\Us/\zeta$.
}
\label{figCompIncomp}
\end{figure*}

In this section we determine the relative importance of the compressible and incompressible parts of the hydrodynamic kinetic energy spectrum for various forcing strengths (as in~\cite{Nore:1997}). 
Given the Bose field $\psi(\textbf{r})=\sqrt{n(\textbf{r})}\exp{[i\phi(\textbf{r})]}$, the flow field is defined as $\textbf{w}(\textbf{r})=\hbar/m\sqrt{n(\textbf{r})}\nabla\phi(\textbf{r})$. The flow field can be Helmholtz-decomposed as $\textbf{w}(\textbf{r})=\textbf{w}^{i}(\textbf{r})+\textbf{w}^{c}(\textbf{r})$, where $\textbf{w}^{i}(\textbf{r})$ is the incompressible part (such that $\nabla\cdot \textbf{w}^i(\textbf{r})=0$) and $\textbf{w}^{c}(\textbf{r})$ is the compressible one (such that $\nabla\times \textbf{w}^c(\textbf{r})=0$). We conveniently perform this decomposition in momentum space, where the Fourier-transformed $\tilde{\textbf{w}}(\textbf{k})$ is split into longitudinal (compressible) and transverse (incompressible) components. The hydrodynamic kinetic energy spectra are then calculated as $E^i(k)\equiv\int m|\tilde{\textbf{w}}^i(\textbf{k})|^2/(2V)\,\delta(|\textbf{k}|-k)\textrm{d}^3\textbf{k}$ and $E^c(k)\equiv\int m|\tilde{\textbf{w}}^c(\textbf{k})|^2/(2V)\,\delta(|\textbf{k}|-k)\textrm{d}^3\textbf{k}$. Those spectra are normalized such that the total compressible $E^c_\mathrm{total}$ and incompressible $E_\mathrm{total}^i$ hydrodynamic kinetic energies are $E^{i,c}_\mathrm{total}=\int E^{i,c}(k)\mathrm{d}k$~\footnote{Note that the hydrodynamic kinetic energy is not equal to the total kinetic energy, as the latter also contains a contribution from density modulations, $\propto |\nabla \sqrt{n(\textbf{r})}|^2$, usually referred to as quantum pressure. Though this component has been discussed before (see, \emph{e.g.}~\cite{Nore:1997,Tanogami:2021}), its role in neither vortex nor wave turbulence is well understood, and we do not focus on this term here.}.

In Fig.~\ref{figCompIncomp}(a) we show an example of $E^{i,c}(k)$ for $a=100a_0$ and $\Us/\zeta=1$. The compressible spectrum appears to be well described by a power law over a certain momentum range, with an exponent $\approx 0.8$. We note that this exponent is not far from what one would get if one were to blindly convert the WWT prediction for $n(k)$ into a prediction for $E^c(k)$: if $E^c(k)\propto k^4n(k)$ with $n(k)\appropto k^{-3}$, then $E^c(k)\appropto k$.

The incompressible-energy spectrum is more complex; intriguingly, around  $k\xi\sim1$, $E^i(k)$ appears power-law-like, with $E^i(k)\propto k^{-1}$, which is reminiscent of claims related to Vinen turbulence~\cite{Cidrim:2017}. We do not find any indication of Kolmogorov's scaling $E^i(k)\propto k^{-5/3}$~\cite{Frisch:1995}; this is not really surprising since Kolmogorov's scaling is expected to hold for $\kF\ll k \ll 1/\ell\ll 1/\xi$, but in our system this condition is never satisfied (here, $\ell$ is the average distance between vortices). 

Fig.~\ref{figCompIncomp}(a) also shows that the compressible energy is larger than the incompressible one in the momentum range $1\lesssim k\xi\lesssim\kD\xi$. To examine this more carefully, we show in Fig.~\ref{figCompIncomp}(b) the ratio $E^{c}(k)/E^{i}(k)$ of the two spectra. 
We see that for all our simulations, $E^c(k)$ is never comfortably dominating $E^{i}(k)$; the situation only gets worse as $\Us/\zeta$ increases (see also~\cite{Fischer:2024}). The fact that the incompressible energy is of the same order as the compressible one might explain why the observed EOS is explained neither by Kolmogorov vortex turbulence nor wave turbulence. 
The shape of the EOS is instead likely due to an interplay between compressible (wave) and incompressible (vortex) excitations. It seems thus particularly remarkable that such an interplay could give rise to such a clean power-law EOS. 

A point of caution is warranted: it is not evident that, despite its widespread use, this Helmholtz decomposition is relevant to understand the EOS, especially in a situation where $E^i$ and $E^c$ are of similar magnitude. 
This decomposition would be most relevant if the momentum-space transport equations for the fluxes were decoupled, however, there is no guarantee of absence of `cross-talk' between the compressible and incompressible energies even in the inertial range. 
In addition, while the energy spectrum is decomposed into two components, we extract only a single $\epsilon$; it might be more appropriate -- if possible -- to consider separate fluxes for the two components.

\end{document}